# High-temperature superconductivity in hydrides: experimental evidence and details.

M. I. Eremets[1]*, V. S. Minkov[1], A. P. Drozdov[1], P. P. Kong[1], V. Ksenofontov[1], S. I. Shylin[2], S. L. Bud'ko[3,4], R. Prozorov[3,4], F. F. Balakirev[5], Dan Sun[6], S. Mozaffari[6,7], L. Balicas[6]

[1]Max Planck Institute for Chemistry; Hahn-Meitner-Weg 1, Mainz 55128, Germany
[2]Department of Chemistry – Ångström Laboratory, Uppsala University, PO Box 523, 75120 Uppsala, Sweden
[3]Ames Laboratory, U.S. Department of Energy, Iowa State University; Ames, IA 50011, United States
[4]Department of Physics and Astronomy, Iowa State University; Ames, IA 50011, United States
[5]NHMFL, Los Alamos National Laboratory, MS E536, Los Alamos, New Mexico 87545, USA
[6]National High Magnetic Field Laboratory, Florida State University, Tallahassee, Florida 32310, USA
[7]Department of Materials Science and Engineering, The University of Tennessee, Knoxville, Tennessee 37996, USA

**Abstract**

Since the discovery of superconductivity at ~200 K in $H_3S$ [1] similar or higher transition temperatures, $T_c$s, have been reported for various hydrogen-rich compounds under ultra-high pressures [2]. Superconductivity was experimentally proved by different methods, including electrical resistance, magnetic susceptibility, optical infrared, and nuclear resonant scattering measurements. The crystal structures of superconducting phases were determined by X-ray diffraction. Numerous electrical transport measurements demonstrate the typical behaviour of a conventional phonon-mediated superconductor: zero resistance below $T_c$, shift of $T_c$ to lower temperatures under external magnetic fields, and pronounced isotope effect. Remarkably, the results are in good agreement with the theoretical predictions, which describe superconductivity in hydrides within the framework of the conventional BCS theory. However, despite this acknowledgement, experimental evidence for the superconducting state in these compounds have recently been treated with criticism [3, 4], which apparently stems from misunderstanding and misinterpretation of complicated experiments performed under very high pressures. Here, we describe in greater detail the experiments revealing high-temperature superconductivity in hydrides under high pressures. We show that the arguments against superconductivity [3, 4] can be either refuted or explained. The experiments on the high-temperature superconductivity in hydrides clearly contradict the theory of hole superconductivity [4] and eliminate it [3].

**Introduction**

The discovery of new superconductors demands strict, unambiguous proof. Obviously, this is applicable to the nearly room-temperature superconductivity (RTSC) discovered in hydrogen sulphide at 203 K at high, megabar-level pressures of ~150 GPa [1, 5] (1 GPa ~ $10^4$ atmospheres). High pressures put significant constraints on the experimental study of superconductivity. At such pressures the typical size of samples residing in a diamond anvil cell (DAC) is necessarily small – of the order of ~50 μm in lateral dimension. Nevertheless, even for DACs there is a number of available experimental methods allowing these tiny samples to be probed using various electrical, magnetic, optical and X-ray diffraction techniques [2, 6]. Electrical transport measurements demonstrated a zero resistance state in $H_3S$ below its $T_c$ [1]. The application of an external magnetic field provided further evidence for superconductivity: $T_c$ shifts to lower temperatures. The isotope effect – a strong decrease of $T_c$ resulting from the replacement of hydrogen atoms by deuterium – clearly indicates the phonon-mediated BCS mechanism of superconductivity [1].

The Meissner effect, or expulsion of magnetic field from a superconductor, is a crucial and independent test for superconductivity. Therefore, samples with $H_3S$ were subsequently probed by high-sensitivity measurements of magnetic susceptibility using a superconducting quantum interference device (SQUID) magnetometer [1] under megabar-pressure conditions. A clear diamagnetic signal (screening of magnetic field) below $T_c$ was observed in excellent agreement with the electrical transport measurements. This result was supported by nuclear resonant scattering measurements [7]. Later the superconducting gap was estimated through infrared spectroscopy [8]. X-ray diffraction gave information on the crystal structure of superconducting phases [9-11].

An essential and crucial step for any newly discovered superconductor is reproducibility. In contrast to the discoveries of $MgB_2$ and the cuprates, where superconductivity was immediately reproduced by many laboratories, the superconductivity in $H_3S$ could not quickly be replicated by other groups because of the complicated synthesis protocol associated with the condensing and loading of $H_2S$ at low temperatures, and the difficulties inherent to measurements at the megabar-pressure range. An important step for the independent confirmation of superconductivity in $H_3S$ was electrical transport measurements of the prepared samples in the laboratory of Prof. K. Shimizu at Osaka University [10]. Shortly afterwards, $H_3S$ was independently synthesized and superconductivity was fully confirmed [12-16]. Later, new hydrides with higher $T_c$s and crystal structures distinct from that of $H_3S$ were discovered by different groups, including $T_c$ ~250minkovminkov K in $LaH_{10}$ [9, 17-19], $T_c$ ~243 K in $YH_9$ [20], and lower $T_c$s in many other metal superhydrides [2].

More recently, room-temperature superconductivity was announced first at 288 K under a pressure of 276 GPa [21] and then at 294 K at 21 GPa [22] in the same sulphur-carbon-hydrogen system. However, these contradictory measurements with no access to the raw data have found serious criticism [23-27]. After more than a year room-temperature superconductivity in the sulphur-carbon-hydrogen system was neither confirmed experimentally nor supported by theoretical calculations. Despite the significant efforts, we also failed to reproduce these observations. Our study on the S-C-H system will be described in detail elsewhere [28].

The experimental evidence for superconductivity in hydrides is consistent with a large number of theoretical predictions and calculations [2] and the general theory of conventional superconductivity [29, 30]. Meanwhile J. Hirsch and F. Marsiglio (H&M) have been trying to explain superconductivity in hydrides through their theory, where superconductivity results from "a few holes conducting through a network of closely spaced negatively charged anions, in conducting substructures with excess negative charge" [4]. In contrast to BCS theory, in the proposed theory [4] superconductivity is driven by an universal mechanism that is not connected with electron-phonon interactions. H&M first tried to explain superconductivity in $H_3S$ [5] by holes, which conduct through the direct overlap of planar $s$–$p$ orbitals of sulphur atoms. Sulphur atoms, instead of hydrogen atoms, was stated to be essential for the "hole superconductivity". Explanation for superconductivity by the hopping holes between negatively charged anions (e.g., sulphur in $H_3S$) obviously fails for metal superhydrides such as $LaH_{10}$ where the rare-earth atoms are cations [9, 17, 31]. The contribution of hydrogen in superconductivity is impossible to ignore because of the isotope effect – substitution of hydrogen atoms by deuterium atoms leads to shift of $T_c$ to lower temperatures – a *negative* shift. The shift of $T_c$ is enormous: $\Delta T_c$ ~70 K for the *P6₃/mmc* phase of $YH_9/YD_9$ [32] and the *Fm-3m* phase of $LaH_{10}/LaD_{10}$. For comparison with the previous studies the largest shift was found in $MgB_2$ with $\Delta T_c$ ~ 1.0 K between $Mg^{11}B_2$ and $Mg^{10}B_2$ [33]. The values of the isotope coefficient in hydrides are close to the value of $\alpha \approx 0.5$ for conventional superconductivity: $\alpha = 0.46$ for the *Fm-3m* phase of $LaH_{10}$ [9], $\alpha = 0.39$ and 0.50 for the *Im-3m* phase of $YH_6$ and *P6₃/mmc* phase of $YH_9$, respectively [32].

An inability to explain the isotope effect clearly shows that the hole superconductivity is invalid. According to this theory, the isotope effect should be small, if any, and with a *positive* shift – $T_c$ should increase for a crystal with heavier atoms. Failed to explain conventional high-temperature superconductivity in hydrides by the hole superconductivity H&M [3, 4] put into doubt the whole evidence of conventional superconductivity: "the experimental observations rule out conventional superconductivity but it could be either unconventional superconductors of a novel kind, or alternatively, that they are not superconductors" [34]. For several years, H&M have been trying to find inconsistencies, contradictions in the published works, demonstrating an apparent misunderstanding of the high-pressure experiments. The arguments against superconductivity raised in Refs [4] can be summarized as follows: *i*) observation of thermal hysteresis in resistance measurements; *ii*) absence of broadening of the superconducting transition under external magnetic fields; *iii*) absence of reliable magnetic measurements.

We address these issues by considering the related experiments in greater detail and including some new results in comparison with the published papers. We hope that this comprehensive analysis of the high-pressure experiments on high-temperature superconductors will be useful to the community.

**Results**

**Thermal hysteresis in temperature-dependent measurements**. The transformation between the normal and superconducting states under zero magnetic field in type-II superconductors is a second order phase transition and, therefore, there should be no hysteresis in the temperature-dependent resistance measurements between cooling and warming cycles. However, typically there is a significant hysteresis in $R(T)$ measurements of hydrides under high pressures. Based on this fact, H&M concluded that most likely, hydrides are not superconductors [35].

Besides the sample property, an imperfect thermal equilibrium between thermometer and sample can also result in extrinsic hysteresis. Upon cooling, the DAC is usually in a cold gas flow (nitrogen or helium), and the temperature of the sample lags with respect to that of the thermometer located in the exterior of the DAC. Obviously, this difference depends on the cooling rate (Fig. 1a–c). Upon slow warming without using a heater (typical rate ~0.3 K min$^{-1}$), the DAC body thermalizes with the cryostat, and consequently the DAC and the thermometer are nearly in equilibrium. Therefore, we and other groups use the $R(T)$ data taken upon warming cycle. We checked this procedure with the DACs whose hysteresis was significantly reduced through a careful insulation of the thermometer from the cooling gas (Fig. 1b). The thermal hysteresis decreased from a value of ~10-20 K to ~3 K, but the $R(T)$ curve upon warming cycle remained nearly the same.

Various cooling/warming protocols are used by the different groups. For instance, in the experiments on LaH$_{10}$ [17] its higher $T_c$ = 260 K was determined from the $R(T)$ data upon cooling. This value could be inaccurate because the cooling rate was not reported. In contrast, in Ref. [19] the transition at $T_c$ = 250 K was determined upon warming cycle and it is consistent with Ref. [9]. Another uncertainty in Ref. [17] is the evaluation of a pressure value, at which the superconducting transition occurred, since it was not determined directly.

The most accurate low-temperature measurements were collected with a miniature DAC (Fig. 1c–f). Since such DACs are diminutive in a size and thermal mass (an outer diameter is 8.8 mm and weight ~ 9 g) and have relatively high thermal conductivity (the DAC body is made of a copper alloy), the thermal gradients are minimized. Still, one observes thermal hysteresis when the temperature sensor is attached to the body of DAC in a typical experimental arrangement (Fig. 1c, e). However, the measurements are much more accurate, if the sensor is attached to the metal gasket, which surrounds the sample compressed between both diamond anvils and, thus, is in a direct thermal contact with it. In this arrangement, the difference in $R(T)$ between cooling and warming cycles is negligibly small (Fig. 1d, f).

**Electrical measurements**. H&M also falsely claim the absence of a broadening of the superconducting transition under external magnetic fields in the $R(T)$ measurements of hydrides, raising doubts about superconductivity in these compounds. Before beginning the discussion on this issue, we present details of electrical transport measurements, which are essential for the interpretation of superconductivity.

A desired superconducting sample is synthesized within the pressure range of its stability – most high-$T_c$ hydrides require pressures over one megabar. For that, the DAC containing a piece of a precursor compound (e.g., chemical elements (S, La, Y) or their stable hydrides ($LaH_3$, $YH_3$)), is placed in a gas loader where it is clamped under hydrogen atmosphere at elevated pressures of 0.10–0.15 GPa. The sample is further pressurized to the required pressure value, at which it transforms into hydrogen-rich product via the diffusion of hydrogen into the precursor compound. The diffusion is a long process at room temperature, but it can be greatly accelerated by heating a sample in an oven (up to hundreds of degrees), or by a pulsed laser (up to thousands of degrees). This method has two important advantages: it yields solely binary products without impurities of third elements, and allows for a good control of the hydrogenation. Excess hydrogen can be easily monitored both optically and spectroscopically when hydrogen surrounds the final product. A much easier but less controlled method is the use of hydrogen containing materials such as $BH_3NH_3$ [17], $AlH_3$, or paraffins [36] as an alternative source of hydrogen. Upon heating, these compounds decompose and release free hydrogen. This technique was recently successfully introduced for the synthesis and studies of lanthanum superhydride by Somayazulu *et al.* [17]. We note, that a long time ago this method was used by Ponyatovskii for the synthesis of hydrides in large-volume high-pressure cells [6, 37]. However, this method has its flaws. For instance, the decomposition products may contaminate the superconducting phase or even react with it. It is also difficult to control the amount of the released hydrogen, which is crucial for the reliable synthesis of the stoichiometric hydrogen-enriched phases that require excess hydrogen.

The electrical resistance is typically measured using a four-contact arrangement. This guarantees that the resistance of the electrical leads is excluded from the measurements – only the sample is examined (Fig. 2). The simplest four-probe arrangement for micrometer-size samples clamped between diamond anvils is the van der Pauw geometry [38, 39] (Fig. 2a, b). By alternating the combination of current and voltage contacts, one can measure resistance from different parts of the sample that provides information about its uniformity (Fig. 2c, d). In addition, by passing electrical current through opposite contacts under an applied magnetic field, one can pick up the Hall signal from another pair of contacts [40].

The interpretation in terms of the emergence of superconductivity when the *R(T)* dependence sharply drops to zero below $T_c$ is obvious for perfect samples (Fig. 2a). However, many real samples are contaminated by unreacted precursor compounds, contain impurity by-product phases (e.g., unsaturated lower hydrides) or are poorly crystallized. These imperfections are often unavoidable, since some tiny areas of a sample are not thoroughly heated in order to prevent the electrical leads from damage by a pulsed laser. In addition, samples of a larger size, which are not surrounded by a quasi-hydrostatic medium (e.g., excess hydrogen) have considerable pressure gradients. In such samples, the superconducting transition broadens and displays additional steps, indicating that the *R(T)* dependence is affected by non-uniform current flow (Fig. 2c–e). These distortions of the superconducting transition were also observed in other superconductors [41]. These interesting multiple resistance steps around the main transition were not studied theoretically. An empirical model for $MgB_2$ [41] attributes these transitions to inhomogeneous doping (resulting in parts of the sample with different $T_c$ values). The shape of *R(T)* is explained by a core-shell model: the outer layers or the shell of grains would correspond to a doped $MgB_2$ phase, which displays the lower $T_c$ while the core of the grains still contains pure $MgB_2$ phase exhibiting the superconducting transition at higher temperatures.

Perhaps even more complicated is another kind of distortions of the superconducting transition: a peak in *R(T)* [42-45] – an anomalous increase of the resistance, that precedes the sharp drop to zero (see Fig. 2c, d). This peak has been observed in many disordered superconductors [46] and has been the subject of many studies [47]. It was systematically studied, for instance, in boron-doped diamond [45, 48]. It was suggested to be due to the granular character of superconductivity: Cooper pairs first condense, and then localize into "bosonic islands", causing a sharp increase in the resistance as the islands are separated from each other [48]. At lower temperatures, percolation through these size-increasing islands leads to the drop in resistance down to zero. This model describes well the peak that is observed in hydrides (see Fig. 2c, d and, for instance, Ref. [49]). We repeatedly observed a peak in $R(T)$ with a particular combination of van der Pauw contacts, whereas a normal sharp step is observed with another combination (Fig. 2c, d). Most likely, this reflects the non-uniformity of the superconducting phase in samples. It is clear that $T_c$ should be defined as the temperature where the peak starts to arise upon cooling (Fig. 2c and Ref [49]), not as the temperature associated to the maximum of the peak. This anomalous peak in $R(T)$ loses amplitude, broadening into a bump under applied magnetic fields [45, 47].

Importantly, not only the observation of zero resistance strongly supports superconductivity in hydrogen-rich compounds, but also the transition imperfections (broadening, steps, and peaks) discussed above, since these features are common among inhomogeneous superconductors.

**Superconducting transitions at high magnetic fields.** Measurements of the superconducting transition in external magnetic fields independently verify the presence of superconductivity and additionally provide valuable information on the properties of superconductors. Whereas the magnetic field has often a small effect on the resistance of normal metals, a $T_c$ is strongly supressed as the magnetic field increases, and superconductivity can be completely suppressed at fields above an upper critical field, which is defined as $H_{c2}$ for type-II superconductors. This suppression is difficult or impossible to achieve in high-$T_c$ hydrides that have very high values of $H_{c2} \geq 100$ T [9, 40, 49, 50]. Nonetheless, available laboratory magnetic fields in the order of ~10 T allow one to define the slope of $H_{c2}(T)$ near $T_c$ and estimate the value of $H_{c2}(0)$ by extrapolating the values of the $H_{c2}(T)$ to lower temperatures [51]. An accurate evaluation of $H_{c2}$ and a study of $T_c(H)$ up to the record static fields up to 45 T was done at the NHMFL in Tallahassee using the hybrid magnet and up to 65 T in pulsed magnetic fields at the NHMFL in Los Alamos through the use of miniature nonmagnetic DACs. Experiments in pulsed magnetic fields are especially demanding due to the potential heating of the DAC and the sample due to eddy currents associated to fast varying magnetic fields. It turned out that the miniature DAC, which was specially designed for SQUID measurements (Fig. 3) is also suitable for magnetoresistance measurements under strong pulsed magnetic fields [40, 52].

We often observe that the steps in $R(T)$, which accompany the superconducting transition in imperfect samples, are strongly affected by the external magnetic field; e.g. they disappear above a modest field of just 3 T in $H_3S$ [40] (Fig. SM1 in Ref [40]). Similar behaviour was also observed in $LaH_{10}$ [9]. The distortions of the superconducting transition are suppressed under high magnetic fields in hydrogen-rich compounds as well as $MgB_2$ [9, 40, 41], see Fig. 2e and also Fig. 2a in Ref [9]. There are no systematic studies of these phenomena, however, we attributed them to inhomogeneities of the superconducting phase in samples, where superconductivity first sets across weak inter-grain links but is easily suppressed by magnetic field [40].

We note that despite the complicated shape of the $R(T)$ dependence in non-uniform samples (bumps, steps, and broadening), the value of a $T_c$ – the temperature corresponding to the emergence of superconductivity – remains nearly unchanged and is close to the value measured in perfect samples.

**Broadening of the superconducting transition under applied magnetic fields.** Previously, we discussed the different reasons for the broadening of superconducting transition: impurities, poor crystallinity, pressure gradients, inhomogeneity of the superconducting phase in samples, etc. A special non-trivial case is the broadening of a superconducting transition in perfect type-II superconductors under strong magnetic fields. Magnetic field penetrates the type-II superconductor in the form of vortices, and vortex motion associated to the electrical currents produces dissipation. As a result, the width of the resistive transition $\Delta T$ is expected to increase with the field as $\sim B^{2/3}$ (Ref. [53]). Indeed, we observed the broadening of the superconducting transitions under magnetic fields (Fig. 4). Naturally, the broadening can be observed only if the superconducting transition is sharp enough in zero-field. Also, one should keep in mind that the vortex dissipation that leads to the broadening of the transition depends on vortex pinning mechanisms and vortex phase diagram. For instance, in cuprates the existence of a vortex liquid phase leads to a dramatic broadening of the superconducting transition under an external magnetic field [54] whereas the smaller broadening was observed for iron-based superconductors [55].

However, in Ref. [3] it is argued that the high-temperature superconducting hydrides display no broadening of the superconducting transition under magnetic fields. Unfortunately, the authors manipulated the published data and cherry-picked the references that include samples exhibiting broad superconducting transitions already in zero-field. For these samples, it is difficult to detect the magnetic field-induced broadening. H&M ignored the publications, which contradict their beliefs, i.e. the initially sharper superconducting transitions that broaden in applied fields, e.g. Ref. [40]. Therefore, the arguments of H&M in Ref. [3] against superconductivity in hydrides, which are largely based on their claims on nonstandard absence of the broadening of the superconducting transitions in external magnetic fields are false. The broadening is indeed observed and is consistent with the well-established scenario where the hydrides are type-II superconductors [1].

There is an apparent exception: Snider *et al.* [21] reported no broadening of the superconducting transition under magnetic fields in their experiments (Fig. 4). This is particularly puzzling because the transitions are extremely sharp $\Delta T < 0.5$ K ($\Delta T/T_c < 0.2\%$), even narrower than those in MgB$_2$ where the broadening is pronounced [41, 56]. Unfortunately, H&M [34] groundlessly extrapolated this controversial result to all other hydrides.

**Magnetic susceptibility measurements: observation of the magnetic field screening.** A major claim of H&M is the "absence of magnetic evidence for superconductivity in hydrides under high pressure". The authors question the measurements of the magnetic moment of H$_3$S samples as a function of both temperature $M(T)$ and magnetic field $M(H)$ [1]. Here, we will elucidate these experiments.

Magnetic susceptibility measurements under very high pressures are very challenging, and in fact much more difficult than the electrical transport measurements. The diameter of a superconducting sample at megabar pressures is smaller than ~100 μm and its thickness is just a few μm. Such small superconducting disk has a magnetic moment of just ~$10^{-6}$–$10^{-7}$ emu that is quite small, and in fact comparable to the sensitivity threshold of a SQUID magnetometer (~$10^{-8}$ emu). Nevertheless, it was possible to perform such measurements even at megabar pressures using the miniature DAC specially designed for this purpose (Fig. 3). The body of the DAC is made of a high-purity Cu-Ti alloy with 3 wt % Ti in order to minimize the magnetic signal over a wide temperature range and provide a high mechanical strength. The piston and diamond seats, which are subjected to the highest load, were made of the harder Cu-Be alloy with 1.8-2.0 wt % Be. Such combination of materials allows us to construct the miniature DAC with an outer diameter of 8.8 mm, which is capable of reaching pressures as high as 220 GPa retaining low overall magnetic response.

So far, measurements of the magnetic susceptibility in a SQUID magnetometer were reported only in the original work [1], for both H$_3$S and D$_3$S (see Fig. 5a, b and Supplementary Fig. 6d in Ref. [1]). A sharp step was observed in the magnetization curve $M(T)$ under zero-field cooled, ZFC, condition at 203 K for H$_3$S and at 136 K for D$_3$S counterpart. It coincides with the superconducting step observed in the $R(T)$ data for both H$_3$S and D$_3$S, and thus provides clear evidence for superconductivity in both compounds. In H$_3$S ZFC and field-cooled, FC, $M(T)$ plots are different – the ZFC curve displays a sharp step, but practically no kink is observed under FC conditions, as the magnetic flux penetrates the material (Fig. 5a). However, some samples still display a small signature for the superconducting transition when measured under FC conditions (Fig. 5b).

The major obstacle for proper magnetic susceptibility measurements is a background from the DAC, which is still much larger than the signal from the sample. It was not possible to subtract the background of the DAC in the usual way by measuring the DAC with and without the sample. Instead, the magnetic susceptibility was measured just above $T_c$ where the sample is in its normal metallic state [1]. This magnetic signal was used as a background for subsequent measurements in SQUID magnetometry. Fortunately, at high temperatures in the vicinity of 200 K the magnetic susceptibility of the materials of the DAC body display a weak dependence on temperature. Apparently, subtraction of a background measured at a single temperature point should not be universal in the whole temperature range below $T_c$, especially at low temperatures. Nevertheless, this approach allowed us to observe the superconducting transition in the $M(T)$ measurements. The absolute value of the diamagnetic transition of ~$10^{-6}$ emu (Fig. 5a) corresponds to the signal from a thin diamagnetic disk with a diameter of ~100 μm – which is comprable to the sample size estimated by optical microscopy. Nevertheless, the subtraction of the background was not perfect and therefore the $M(T)$ curve is shifted vertically due to a paramagnetic component (Fig. 5a), and this was criticized in Ref. [57]. However, this superconducting transition fully agrees with the recent and more accurate measurements [58] (Fig. 5c–h).

Measurements of hysteresis in $M(H)$ reported in Ref. [1] are even more difficult because the magnetic susceptibility of the materials of the DAC body, as well as copper wires, solder and electrical leads mounted for the corroborative resistance measurements, depends on a magnetic field as well. A lower critical magnetic field $H_{c1}$ of ~30 mT was only roughly estimated as the point associated to the bending of the $M(H)$ hysteretic loops without considering the geometry of the sample (demagnetization factor) (Fig. 4c, d in Ref [1]). However, the correct value of the inflection point in the $M(H)$ virgin curves was determined in the recent work [58] and its value is higher ~96 mT at 0 K for $H_3S$. This is not a large disagreement keeping in mind the absence of $M(H)$ virgin curves in Ref. [1] and related uncertainties associated with the criterion of the deviation of $M(H)$ from linearity in hysteretic loops. Even for the less experimentally challenging magnetic measurements at ambient pressure conditions, one can see factor 2-3 difference among different publications for the penetration depth (that results in even larger spread of the $H_{c1}$ values) evaluated for the same superconductors [59].

The original magnetic measurements [1] were criticized in Ref [34] and even qualified as a "myth" [24] given that they had not been reproduced for six years. In fact, we have a number of unpublished results, which agree with the Ref. [1] (Fig. 5b). Here, we present one of such results in Fig. 6, which was obtained right after the publication of Ref. [1]. We attempted to evaluate the lower critical field $H_{c1}$ using the virgin curve of $M(H)$. The sample was prepared according to the original recipe, i.e. through the pressure-induced disproportionation of compressed $H_2S$, and subsequently pressurized to $P$ ~140 GPa. We did not characterize the sample by X-ray diffraction, however, the $T_c$ value of ~140 K, which was observed in the ZFC $M(T)$ measurements (Fig. 6a) suggests that the sample had the rhombohedral distorted crystal lattice or the $R3m$ phase of $H_3S$, according to the later established dependence $T_c(P)$ (Fig. 3 in Ref. [11]). Fig. 6 shows magnetization measurements of the $H_3S$ sample at $T$ = 100 K. The measurements were performed in the background subtraction mode using the reference data measured from the same DAC at $T$ = 210 K. Original hysteretic loop shown in the inset in Fig. 6b reveals a strong paramagnetic component originating mainly from the DAC. After subtraction of this paramagnetic background, which is approximated by linear dependence $M(H)$, the superconducting hysteresis has the typical shape for a type-II superconductor (Fig. 6b). At $H$ = 30 mT the $M(H)$ virgin curve starts to deviate from a straight line allowing the determination of a lower critical field $H_{c1}$. Taking into account a demagnetization correction $\frac{1}{1-N}$ ~9 for a thin disk shaped sample, which was estimated from the measured value of $\Delta M$, we obtain $H_{c1}$ (100 K) = 270 mT. The extrapolation of $H_{c1}$ to lower temperatures using the equation $H_{c1}(T) = H_{c1}(0\ K) \cdot (1 - \left(\frac{T}{T_c}\right)^2)$ yields the value $H_{c1}$(0 K) ~550 mT.

The ZFC $M(H)$ dependence coincides with the full magnetization loop at $H \approx 0.2$ T. Considering also the demagnetization factor one can estimate the full penetration field $H_p(100)$ K $\approx 1.8$ T. Note, that the hysteretic loop has an asymmetric shape and is broader than that in Ref. [1]. The hysteresis asymmetry results from the polycrystalline nature of the sample. According to Refs [60, 61] the pronounced asymmetry is observed when the grain size $R < 10\lambda_L$ ($\lambda_L$ is the London penetration depth). One can expect more symmetric hysteretic curves for samples of better quality with larger grain size. The dependence of the superconducting hysteresis on the grain size was systematically studied in $MgB_2$ in Refs [56, 62].

For a long time we did not clearly appreciate the role played the quality and integrity of the samples. Even when the sample has a large enough size and clearly shows superconductivity in electrical transport measurements, the magnetic susceptibility signal can turn out to be elusive or below the sensitivity of the SQUID magnetometer. The reason for that can be the granular or non-uniform distribution of the superconducting phase in samples. The electrical current finds a continuous path through superconducting grains and metallic grain boundaries in the transport measurements whereas much smaller thin superconducting grains have a relatively small superconducting volume leading to a smaller signal due to the demagnetization factor.

Only recently, we succeeded in improving our measurements of the magnetic susceptibility significantly [58] (see, Fig. 5c–h). For that we used a solid $BH_3NH_3$ as an alternative source of hydrogen instead of pure hydrogen. This allowed us to produce large samples with a diameter close to that of the culet of the diamond anvils, and to obtain a pronounced diamagnetic signal from superconducting phases under high pressures. We believe that this approach can be adopted by other laboratories.

We were able to extract the small diamagnetic signal of a superconductor from the measured overall magnetic moment, which includes magnetic moments of the DAC body, diamonds and rhenium gasket, using a different approach. First, we measured the magnetic signal of DACs with the starting pressurized precursors (S or $LaH_3$ and $BH_3NH_3$) before laser heating. Then we subtracted these reference data from the magnetic moment collected from the same DACs after laser heating and chemical synthesis and obtained the superconducting response of the $H_3S$ and $LaH_{10}$ compounds. The temperature dependence of the magnetic moment of both samples changed dramatically after laser heating, indicating that they became superconducting (Fig. 5c–h). We note, that Ref [63] incorrectly states that ZFC and FC curves for $H_3S$ coincide at 200 K before heating and after heating, which is clearly not the case for Figs 5d,f. Abrupt steps in the ZFC $M(T)$ curves yield values of $T_c \sim 231$ K for $Fm$-$3m$-$LaH_{10}$ at 130(8) GPa and $\sim 196$ K for $Im$-$3m$-$H_3S$ at 155(5) GPa. These values are in excellent agreement with the previously reported values from four-probe electrical transport measurements under the same pressures[1, 11, 15, 40]. FC curves show the small step for the $Im$-$3m$-$H_3S$ sample (Figs 5h) and the subtle step for the $Fm$-$3m$-$LaH_{10}$ samples. The subtle steps observed on FC curves or their apparent absence is common for superconductors with strong pinning of vortices [64].

In addition, we determined lower critical fields and consequently the London penetration depth for both $H_3S$ and $LaH_{10}$ [58]. For $H_3S$, a lower critical field $H_{c1}(0$ K$)$ is estimated to be $\sim 0.74$–$1.09$ T, leading to the London penetration depth $\lambda_L(0$ K$) \sim 18$–$23$ nm. For $LaH_{10}$, the lower critical field $H_{c1}(0$ K$)$ is estimated to be $\sim 0.42$–$1.75$ T and $\lambda_L(0$ K$) \sim 14$–$35$ nm. The coherence length, $\xi(0$ K$)$, was estimated from our previous measurements of the dependence of $T_c$ on external magnetic fields: it is 1.84 nm for $H_3S$ [40] and 1.51 nm for $LaH_{10}$ [9]. These values yield the Ginzburg-Landau parameter $\kappa = \frac{\lambda}{\xi} \sim 10$–$13$ for $H_3S$ and $\sim 10$–$23$ for $LaH_{10}$. These parameters indicate that both $H_3S$ and $LaH_{10}$ compounds belong to the category of "moderate" type-II superconductors, rather than being hard superconductors as would be intuitively expected from their high $T_c$s. One can see that the coherence length is comparable to those of other high-temperature superconductors, e.g. the cuprates, but the London penetration depth is significantly shorter [65].

**Critical current density.** Using the values of the thermodynamic critical field, $H_c(0$ K$) = 5.6$ T for $H_3S$ and 5.1 T for $LaH_{10}$, and the London penetration depth, $\lambda_L(0$ K$)$ we can evaluate the depairing critical current density as $j_d \sim H_c/\mu_0\lambda$. This yields values of $j_d \sim 1.4 \times 10^{10}$ A cm$^{-2}$ and $\sim 7.4 \times 10^9$ A cm$^{-2}$ for $H_3S$ and $LaH_{10}$, respectively, which are approximately two orders of magnitude higher than the $j_d(0$ K$)$ extracted for $YBa_2Cu_3O_{7-\delta}$ [66] and three orders of magnitude higher than those of $MgB_2$ [67] or $V_3Si$ [68]. It has to be mentioned that the depairing critical current density represents the theoretical maximum or the critical current density and is usually found to be of nearly one order of magnitude higher than the actual depinning critical current density in a number of materials.

The experimentally evaluated critical current densities from irreversible magnetization measurements are significantly lower values $\sim 7 \times 10^6$ A cm$^{-2}$ at 100 K for $H_3S$ and $LaH_{10}$ [58]. We note that the recent critisim of the estimate incorrectly ignores substantial demagnetization factors [63]. The electrical transport current-voltage measurements, yield similar order of magnitude values, viz $\sim 2.6 \times 10^7$ A cm$^{-2}$ for $YH_6$ [49] and $\sim 2.7 \times 10^6$ A cm$^{-2}$ at 4.2 K for $La(Y)H_{10}$ [69]. This could suggest that there is room for improvement of the critical currents in the hydrides.

**Magnetic field screening in the nuclear resonance scattering experiments [7]. Numerical analysis.** Further evidence for superconductivity was demonstrated on the screening of the magnetic field by $H_2S$ compressed under 153 GPa in Ref [7]. In this work, a thin $^{119}$Sn film was inserted inside the $H_3S$ sample, and the magnetic field in the film was monitored via nuclear resonance scattering of synchrotron radiation. In that experiment, an external magnetic field of about 0.7 tesla was shielded by a $H_3S$ superconductor at least up to 140 K. Importantly, the same field was attenuated only to half of this value when it was applied along the sample. Considering nearly zero demagnetization factor in this configuration, it cannot be understood if the sample around Sn foil was not superconducting. This experiment was questioned in Refs [4-6, 8] where the authors claim that the magnetic field should not have been expelled. We note that magnetic field expulsion (i.e. in a FC experiment) is very different from the magnetic field screening (i.e. a ZFC experiment when magnetic field is applied at low temperature after cooling in zero field.).

The ability to shield a magnetic field by a type-II superconductor involves Meissner-London screening at the sample edge [70] and penetration of Abrikosov´s vortices forming a Bean critical state [71]. Unlike any other magnetic material, the latter is non-local on the scale of the whole sample. We therefore performed the measurements in two orientations of an applied field with respect to the sample assembly. A careful analysis of both phenomena [72] in the experimental geometry of [15] suggests that the average magnetic field measured in the interior cavity of a superconductor in two orientations is fully consistent with the conventional behaviour of a type-II superconductor with a lower critical field, $0.3 \leq H_{c1} \leq 0.6$ T and critical current density larger than $j_c \sim 4 \times 10^6$ A cm$^{-2}$. Considering large uncertainties of the estimates, these values are in a good agreement with the estimates obtained from the magnetization measurements discussed above. On the other hand, the difference found in two orientations cannot be understood assuming any other non-superconducting state of the studied sample.

Based on a detailed analysis of the experiment discussed in Ref [7], it appears that the nuclear resonance scattering is a new, non-trivial, and sophisticated technique to detect superconductivity. Hopefully it will find further use to study novel near room-temperature superconductors in difficult conditions, such as ultra-high pressure, and, perhaps in compounds exhibiting superconductivity even above room temperature. A full analysis of this experiment is given elsewhere [72].

**Conclusions.** To conclude, we presented a detailed description of experiments such as the observation of zero resistance that is concomitant with the onset of the pronounced diamagnetic response, as well as the broadening of the resistive transition under external applied fields. The observed temperature – magnetic-field phase diagram is akin to those observed in orbitally limited superconductors. These observations represent standard as well as very solid evidence for high-temperature superconductivity in hydrogen-rich compounds under high pressures and disavow the doubts raised in Refs [3, 4]. The pronounced isotope effect in hydrides is in favour of conventional superconductivity, but contradicts the hole superconducting mechanism [4], and in the essence eliminate this theory, as this declared in Ref [3].

**Author Contributions**

The group in Mainz (M.I.E., V.S.M., A.P.D., P.P.K.) synthesized samples and performed electrical transport measurements; F.F.B., D.S., S.M., and L.B. performed measurements of electrical resistance under high magnetic fields; V.S.M, V.K. and S.I.S. performed magnetic measurements in SQUID; F.F.B., S.L.B., V.S.M. and V.K. analyzed magnetization data; R.P. contributed finite element modelling. All authors collaborated in the discussion of the manuscript. M.I.E. wrote the manuscript along with V.S.M., F.F.B., S.L.B., V.K. and R.P.

**Acknowledgements**

M.E. is thankful to the Max Planck community for valuable support and U. Pöschl for the encouragement. The authors thank M. Tkacz, V. B. Prakapenka, M. A. Kuzovnikov, S. P. Besedin and D. A. Knyazev for collaboration, and I. A. Troyan and I. S. Lyubutin and for useful comments. The authors appreciate the support and collaboration of K. Shimizu and M. Einaga. Work at Ames Laboratory (SLB, RP) was supported by the U.S. Department of Energy, Office of Science, Basic Energy Sciences, Materials Sciences and Engineering Division. Ames Laboratory is operated for the U.S. Department of Energy by Iowa State University under Contract No. DE-AC02-07CH11358. The National High Magnetic Field Laboratory is supported by the National Science Foundation through NSF/DMR-1644779, the State of Florida, and the U.S. Department of Energy.

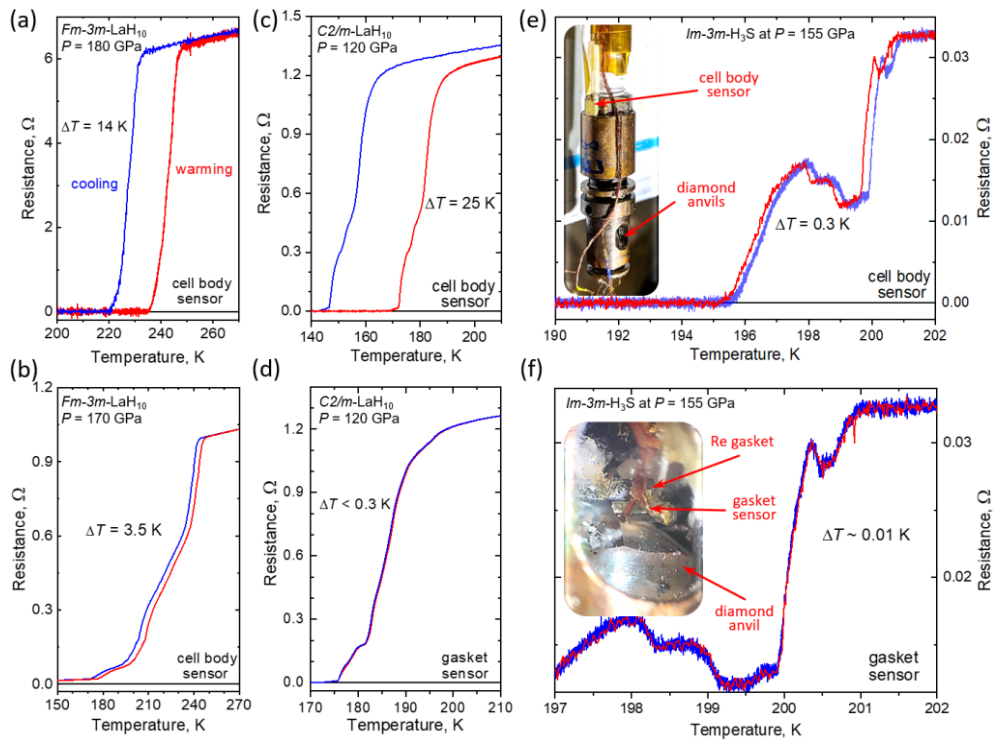

**Figure 1**. The superconducting transitions in LaH$_{10}$ and H$_3$S compounds detected through the temperature dependence of the electrical resistance upon cooling (blue curves) and warming (red curves). (**a**, **b**) *R(T)* measurements of the *Fm-3m*-LaH$_{10}$ phase residing in a large DAC with an outer diameter of 33 mm. The temperature sweep d*T*/d*t* is ~1 K min$^{-1}$ upon cooling and ~0.3 K min$^{-1}$ upon warming. (**a**) The temperature sensor is attached to the body of DAC, leading to a hysteresis of ~14 K. (**b**) The sensor is attached to the body of DAC and carefully insulated from the cooling gas. In this case, the hysteresis is significantly lower ~3.5 K. (**c**, **d**) *R(T)* measurements of the *C2/m*-LaH$_{10}$ phase in the miniature DAC with an outer diameter of 8.8 mm. (**c**) The sensor is attached to the body of the DAC and d*T*/d*t* is ~3 K min$^{-1}$ upon cooling and ~0.3 K min$^{-1}$ upon warming, leading to a hysteresis of ~25 K. (**d**) The same DAC but the sensor is attached to the gasket and d*T*/d*t* is ~0.15 K min$^{-1}$ upon cooling and warming, leading to a hysteresis of ~0.3 K. (**e**, **f**) Superconducting transitions in the *Im-3m*-H$_3$S phase residing in the miniature DAC with d*T*/d*t* of 0.15-1.5 K min$^{-1}$. (**e**) The sensor is attached to the body of DAC and the hysteresis is ~0.3 K; (**d**) the sensor is attached to the gasket and the hysteresis is ~0.01 K. Insets: photos of the miniature DAC showing the location of the temperature sensors.

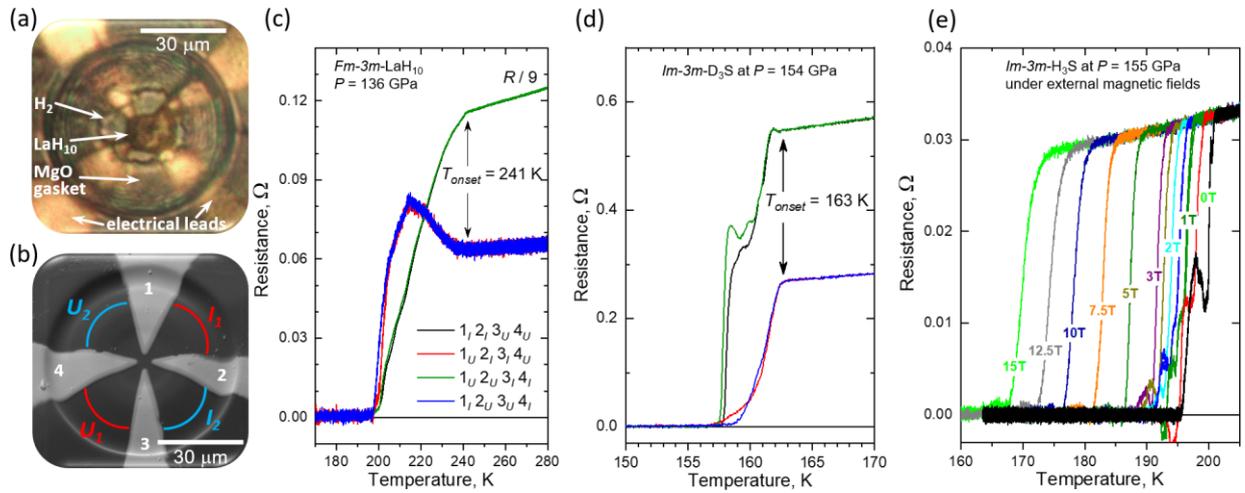

**Figure 2**. Typical four-probe electrical transport measurements in DACs. (**a**) Photo of the sample of *Fm-3m*-LaH$_{10}$ phase with exceess H$_2$ at 138 GPa. (**b**) SEM photo of four Ta/Au electrical leads spattered on the diamond culet before loading the sample. As an example, two out of four possible combinations of current and voltage contacts, which can be realized in the typical *R(T)* measurements are marked by blue and red. (**c**–**f**) *R(T)* measurements displaying the distortions of superconducting transitions (broadening, steps and peaks) in the non-uniform samples. (**c**) Superconducting transition in *Fm-3m*-LaH$_{10}$ at 136 GPa: two combiations (green and black) show the broadened transitions with a $T_{onset}$ of ~ 241 K, whereas other two combinations (red and blue) show a peak in *R(T)*, that precedes the sharp drop to zero-resistance. (**e**) The broadening of the superconducting transition (red and blue) and additional steps in *R(T)* (black and green), indicating that the *R(T)* is affected by non-uniform current flow through the sample of *Im-3m*-D$_3$S at 154 GPa. (**d**) The supression of distortions and bumps at the superconducting transition under external magnetic fields in the *Im-3m*-H$_3$S phase at 155 GPa (taken from Ref [40]).

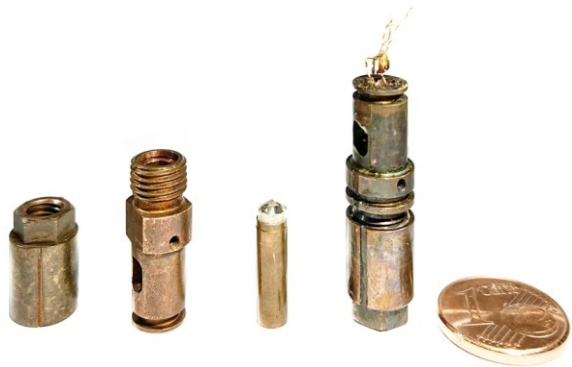

**Figure 3**. The design of the miniature nonmagnetic DAC. A photo of the assembled miniature DAC (right) and its parts (left).

(a)                                                                                     (b)

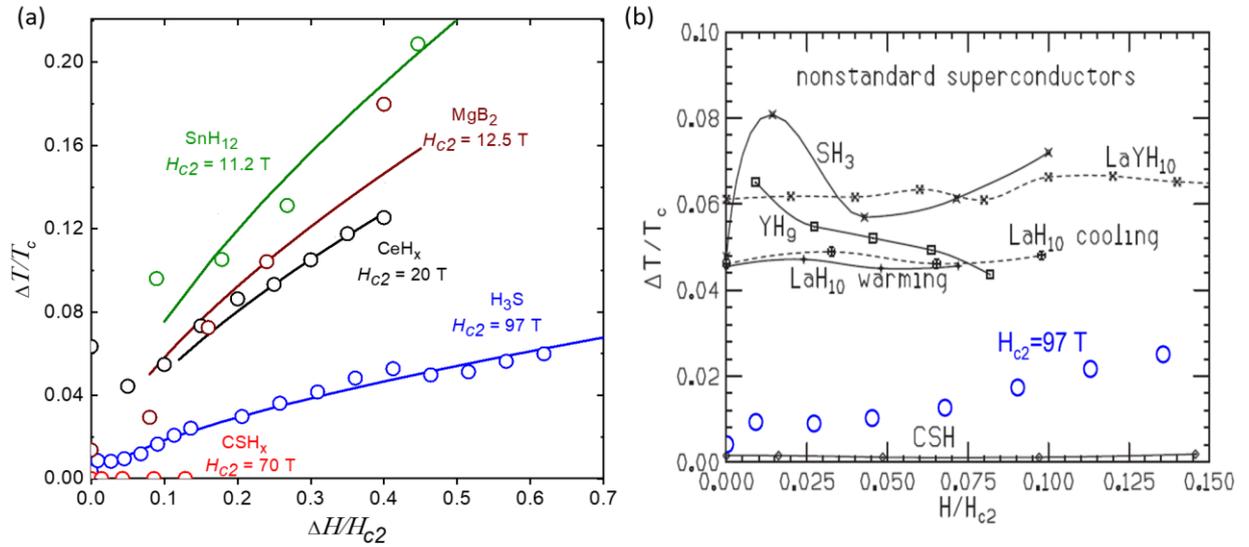

**Figure 4.** Broadening of the superconducting transition under external magnetic fields in different superconducting compounds: (a) $SnH_{12}$ [73], $MgB_2$ [56], $CeH_x$ [74], $H_3S$ [40], C-S-H [21]. The related coordinates are chosen for easier comparison. The solid lines are a guide to the eye for $\Delta T(H) \sim H^{2/3}$ broathening expected at high fields [55]. (b) Broadening of the superconducting transition derived in Ref. [34] is compared with the present study (blue points).

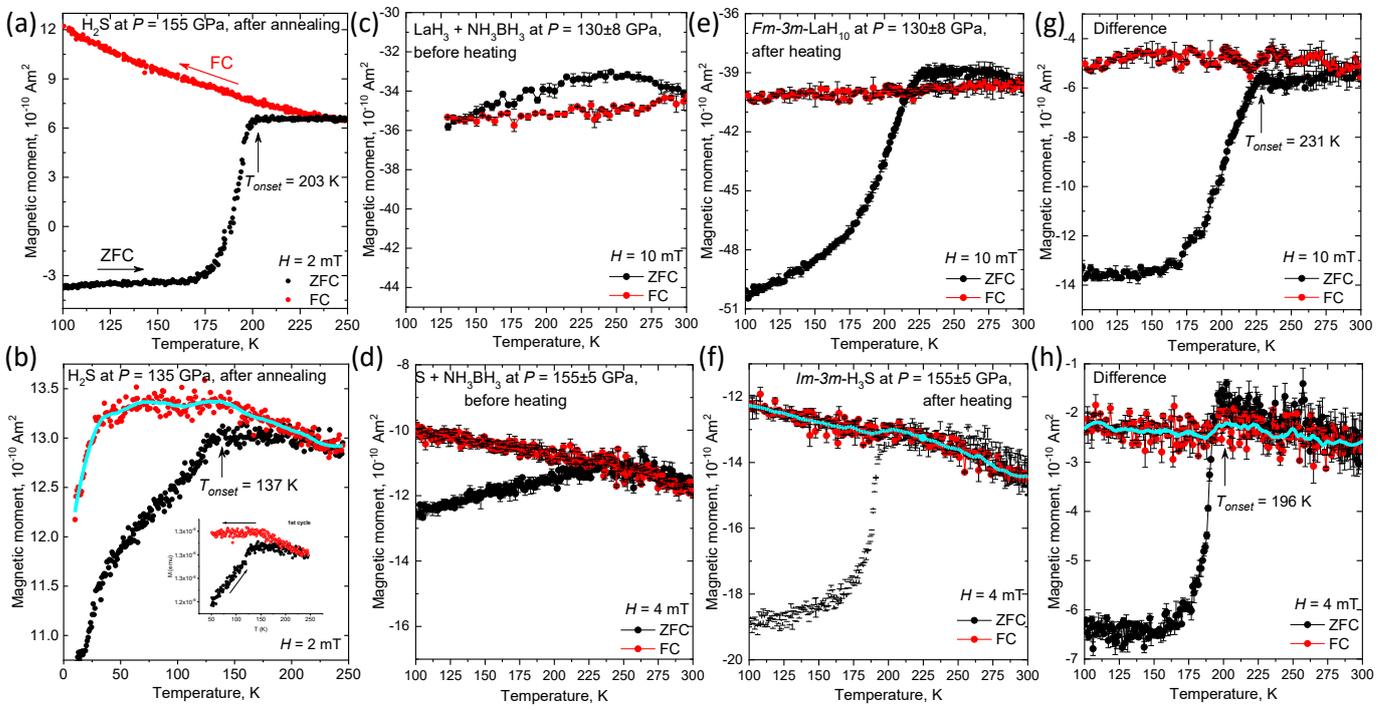

**Figure 5**. Magnetization measurements under high pressures using the miniature DAC. (**a**) $M(T)$ measurements of $H_3S$ at 155 GPa (taken from Ref [1]). (**b**) More typical, the smaller magnetic signal is registered from the $H_3S$ sample at a pressure of 135 GPa. Light and dark blue lines represent smoothed FC and ZFC data. The measurements of the same sample in another run are shown in the inset. (**c**–**h**) The recent improved measurements of $M(T)$ under high pressure with accurate background subtraction (taken from Ref [58]). (**c**, **e**, **g**) $M(T)$ data of the $Fm\text{-}3m\text{-}LaH_{10}$ at 130±8 GPa under magnetic field $H = 10$ mT: (**c**) $M(T)$ of the sandwiched sample with $LaH_3$ and $NH_3BH_3$ before laser heating; (**e**) the heated sample with the superconducting $Fm\text{-}3m\text{-}LaH_{10}$ phase; and (**g**) the difference plot. (**d**, **f**, **h**) $M(T)$ data of the $Im\text{-}3m\text{-}H_3S$ at 155±5 GPa under magnetic field $H = 4$ mT: (**d**) $M(T)$ of the sandwiched sample with S and $NH_3BH_3$ before laser heating; (**f**) the heated sample with the superconducting $Im\text{-}3m\text{-}H_3S$ phase; and (**h**) the difference plot. The black and red circles correspond to ZFC and FC measurements.

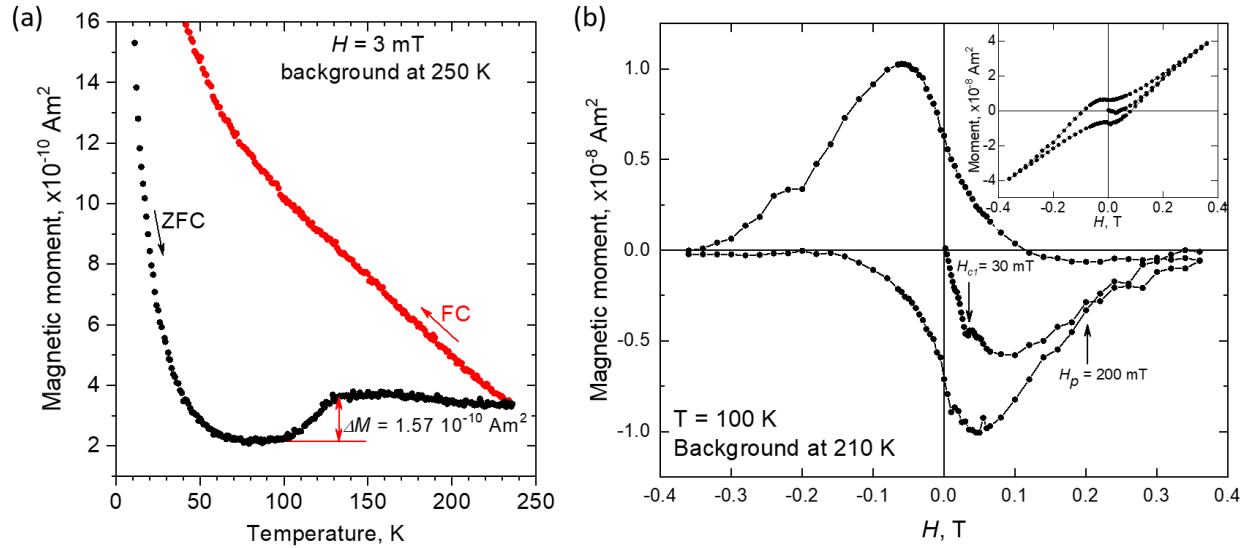

**Figure 6.** Magnetization measurements of superconducting $H_3S$ at pressure $P \sim 140$ GPa. (**a**) $M(T)$ data measured at $H = 3$ mT after subtraction of the background collected at 250 K. (**b**) $M(H)$ data measured at 100 K showing the initial virgin curve and hysteretic loop on alternating the magnetic field. Inset shows the original hysteretic loop before subtraction of the background signal originating from the DAC at 210 K.

**References**

1. Drozdov, A.P., et al., *Conventional superconductivity at 203 K at high pressures.* Nature 2015. **525**: p. 73.
2. Flores-Livas, J.A., et al., *A perspective on conventional high-temperature superconductors at high pressure: Methods and materials.* Physics Reports, 2020. **856**: p. 1-78.
3. Hirsch, J.E., *Hole Superconductivity xOr Hot Hydride Superconductivity.* J. Appl. Phys. , 2021.
4. Hirsch, J.E., *Hole superconductivity.* Phys. Lett. A, 1989. **134**: p. 451.
5. Drozdov, A.P., M.I. Eremets, and I.A. Troyan, *Conventional superconductivity at 190 K at high pressures* arXiv:1412.0460, 2014.
6. Eremets, M.I., *High pressures experimental methods*. 1996, Oxford: Oxford University Press.
7. Troyan, I., et al., *Observation of superconductivity in hydrogen sulfide from nuclear resonant scattering.* Science, 2016. **351**(6279): p. 1303.
8. Capitani, F., et al., *Spectroscopy of H3S: evidence of a new energy scale for superconductivity.* Nature Physics, 2017. **13**: p. 859–863.
9. Drozdov, A.P., et al., *Superconductivity at 250 K in lanthanum hydride under high pressures* Nature, 2019. **569** p. 528.
10. Einaga, M., et al., *Crystal Structure of 200 K-Superconducting Phase of Sulfur Hydride.* Nature Physics, 2016.
11. Minkov, V.S., et al., *Boosted Tc of 166 K in superconducting D3S synthesized from elemental sulfur and hydrogen.* Angew. Chem. Int. Ed,, 2020. **59**: p. 1 – 6.
12. Shimizu, K., et al., *Superconductivity and structural studies of highly compressed hydrogen sulfide.* Physica C: Superconductivity and its applications, 2018. **552** (15): p. 27-29.
13. Shimizu, K., *Investigation of Superconductivity in Hydrogen-rich Systems.* J. Phys. Soc. Jpn., 2020. **89**: p. 051005.
14. Akashi, R., *Evidence of Ideal Superconducting Sulfur Superhydride in a Pressure Cell.* JPSJ News and Comments, 2019. **16** p. 18.
15. Nakao, H., et al., *Superconductivity of Pure H3S Synthesized from Elemental Sulfur and Hydrogen.* J. Phys. Soc. Jpn. , 2019. **88**: p. 123701
16. Huang, X., et al., *High-temperature superconductivity in sulfur hydride evidenced by alternating-current magnetic susceptibility.* National Science Review, 2019. **6**(4): p. 713-718.
17. Somayazulu, M., et al., *Evidence for Superconductivity above 260 K in Lanthanum Superhydride at Megabar Pressures.* Phys. Rev. Lett., 2019. **122** p. 027001.
18. Hong, F., et al., *Superconductivity of Lanthanum Superhydride Investigated Using the Standard Four-Probe Configuration under High Pressures.* Chin. Phys. Lett.. , 2020. **37**: p. 107401.
19. Hong, F., et al., *Superconductivity of Lanthanum Superhydride Investigated Using the Standard Four-Probe Configuration under High Pressures.* Chin. Phys, Lett., 2020. **37** p. 107401.
20. Kong, P., et al., *Superconductivity up to 243 K in the yttrium-hydrogen system under high pressure.* Nature Communications, 2021. **12**(1): p. 5075.
21. Snider, E., et al., *Room-temperature superconductivity in a carbonaceous sulfur hydride.* Nature, 2020. **586**: p. 373–377.
22. Dias, R.P. and A. Salamat, *Standard Superconductivity in Carbonaceous Sulfur Hydride.* arXiv:2111.15017v1 2021.
23. Hirsch, J.E. and F. Marsiglio, *Unusual width of the superconducting transition in a hydride.* Nature, 2021. **596**: p. E9.
24. Hirsch, J.E. and F. Marsiglio, *Absence of magnetic evidence for superconductivity in hydrides under high pressure.* Physica C: Superconductivity and its Applications 2021. **584**: p. 1353866.
25. Dogan, M. and M. L.Cohen, *Anomalous behavior in high-pressure carbonaceous sulfur hydride.* Physica C: Superconductivity and its Applications, 2021. **583**: p. 1353851.
26. Gubler, M., et al., *Missing theoretical evidence for conventional room temperature superconductivity in low enthalpy structures of carbonaceous sulfur hydrides.* arXiv:2109.10019v1, 2021.

27. Wang, T., et al., *Absence of conventional room temperature superconductivity at high pressure in carbon doped H3S.* Phys. Rev. B 2021. **104**: p. 064510.
28. Minkov, V., et al., *Experiments on superconductivity in S-C-H system at high pressure. Raman. X-ray and electrical studies.* To be published. 2021.
29. Bardeen, J., L.N. Cooper, and J.R. Schrieffer, *Theory of Superconductivity.* Phys. Rev., 1957. **108** p. 1175-1204.
30. Ashcroft, N.W., *Hydrogen Dominant Metallic Alloys: High Temperature Superconductors?* Phys. Rev. Lett., 2004. **92**: p. 187002.
31. Peng, F., et al., *Hydrogen Clathrate Structures in Rare Earth Hydrides at High Pressures: Possible Route to Room-Temperature Superconductivity.* Phys. Rev. Lett., 2017. **119** p. 107001.
32. Kong, P.P., et al., *Superconductivity up to 243 K in the yttrium-hydrogen system under high pressure.* Nature Comm., 2021. **12**: p. 5075.
33. Bud'ko, S.L., et al., *Boron Isotope Effect in Superconducting MgB2.* Phys. Rev. Lett., 2001. **86**: p. 1877.
34. Hirsch, J.E. and F. Marsiglio, *Nonstandard superconductivity or no superconductivity in hydrides under high pressure.* Phys, Rev, B 2021. **103** p. 134505.
35. Hirsch, J.E. and F. Marsiglio, *Intrinsic hysteresis in the presumed superconducting transition of hydrides under high pressure.* arXiv:2101.07208v1, 2021.
36. Meier, T., et al., *Proton mobility in metallic copper hydride from high-pressure nuclear magnetic resonance.* Phys. Rev. B, 2020. **102**: p. 165109.
37. E. G. Ponyatovskii, V.E. Antonov, and I.T. Belash, *Properties of high pressure phases in metal-hydrogen systems.* Sov. Phys. Usp., 1982. **25**: p. 596.
38. Pauw, L.J.v.d., *A method of measuring specific resistivity and Hall effect of discs of arbitrary shape.* Philips Research Reports., 1958 **13**: p. 1–9.
39. Pauw, L.J.v.d., *A method of measuring the resistivity and Hall coefficient on lamellae of arbitrary shape.* Philips Technical Review, 1958. **20**: p. 220–224.
40. Mozaffari, S., et al., *Superconducting phase diagram of H3S under high magnetic fields.* Nature Communications, 2019. **10**: p. 2522
41. Wang, C.C., et al., *Superconducting transition width under magnetic field in MgB2 polycrystalline samples.* J. Appl. Phys., 2010. **108**: p. 093907.
42. Troyan, I.A., et al., *Anomalous High-Temperature Superconductivity in YH6.* Adv. Mater., 2021: p. 2006832.
43. Zhang, W., et al., *Unexpected Stable Stoichiometries of Sodium Chlorides.* Science, 2013. **342**: p. 1502.
44. Ems, S.C. and J.C. Swihart, *Resistance peak at the superconducting transition of thin films of tin and indium.* Phys, Lett., 1971. **37A**: p. 255.
45. Zhang, G., et al., *Bosonic Anomalies in Boron-Doped Polycrystalline Diamond.* Phys. Rev. Appl., 2016. **6**: p. 064011.
46. Ems, S.C. and J.C. Swihart, *Resistance peak at the superconducting transition of thin films of tin and indium.* Phys, Lett., 1971. **37A**: p. 255.
47. Crusellas, M.A., J. Fontcuberta, and S. Pinol, *Giant resistive peak close to the superconducting transition in L1-x CexCu04 single crystals.* Phys. Rev. B, 1992. **46**.
48. Zhang, G., et al., *Metal–Bosonic Insulator–Superconductor Transition in Boron-Doped Granular Diamond.* Phys. Rev. Lett, 2013. **110** p. 077001
49. Troyan, I.A., et al., *Anomalous High-Temperature Superconductivity in YH6.* Adv. Mater, 2021: p. 2006832.
50. Kong, P.P., et al., *Superconductivity up to 243 K in yttrium hydrides under high pressure.* arXiv:1909.10482 2019.
51. Werthamer, N.R., E. Helfand, and P.C. Hohenberg, *Temperature and Purity Dependence of the Superconducting Critical Field, $H_{c2}$. III. Electron Spin and Spin-Orbit Effects.* Physical Review, 1966. **147**(1): p. 295-302.

52. Sun, D., et al., *High-temperature superconductivity on the verge of a structural instability in lanthanum superhydride.* Nature Communications, 2021.
53. Tinkham, M., *Resistive Transition of High-Temperature Superconductors.* Phys. Rev. Lett., 1988. **61**: p. 1658.
54. Blatter, G., et al., *Vortices in high-temperature superconductors.* Reviews of Modern Physics, 1994. **66**(4): p. 1125-1388.
55. Vedeneev, S.I., et al., *Temperature dependence of the upper critical field of FeSe single crystals.* Phys. Rev. B 2013. **87**: p. 134512
56. Finnemore, D.K., et al., *Thermodynamic and Transport Properties of Superconducting Mg10B2.* Phys. Rev. Lett., 2001. **86**: p. 2420.
57. Hirsch, J.E. and F. Marsiglio, *Flux trapping in superconducting hydrides under high pressure.* arXiv:2104.03925 2021.
58. Minkov, V.S., et al., *The Meissner effect in high-temperature hydrogen-rich superconductors under high pressure.* Research Square, 2021.
59. Poole Jr, C.P., et al., *9 - Type II superconductivity*, in *Superconductivity (Third Edition)*, C.P.P.P.A.F.J. Creswick, Editor. 2014, Elsevier: London. p. 355-424.
60. D. M. Gokhfeld, et al., *Magnetization asymmetry of type-II superconductors in high magnetic fields.* J. Appl. Phys. , 2011. **109** p. 033904.
61. Gokhfeld, D.M., *An Extended Critical State Model: Asymmetric Magnetization Loops and Field Dependence of the Critical Current of Superconductors* hysics of the Solid State, 2014. **56**: p. 2380–2386.
62. Varghese, N., et al., Journal of Alloys and Compounds, 2009. **470**: p. 63–66.
63. Hirsch, J.E. and F. Marsiglio, *Clear evidence against superconductivity in hydrides under high pressure.* arXiv:2110.07568v1, 2021.
64. Hosono, H., et al., *Exploration of new superconductors and functional materials, and fabrication of superconducting tapes and wires of iron pnictides.* Sci. Technol. Adv. Mater. , 2015. **16** p. 033503.
65. Poole Jr, C.P., et al., *1 - Properties of the normal state*, in *Superconductivity (Third Edition)*, C.P.P.P.A.F.J. Creswick, Editor. 2014, Elsevier: London. p. 1-31.
66. Kunchur, M.N., et al., *Pair-Breaking Effect of High Current Densities on the Superconducting Transition in YBa2Cu3O7-d* Phys. Rev. Lett. , 1994. **75**: p. 752.
67. Kunchur, M.N., S.-I. Lee, and W.N. Kang, *Pair- breaking critical current density of magnesium diboride.* Phys. Rev. B 2003. **68**: p. 064516.
68. Zehetmayer, M. and J. Hecher, *Testing V3Si for two-band superconductivity.* Supercond. Sci. Technol. , 2014. **27**: p. 044006.
69. D. V. Semenok, et al., *Superconductivity at 253K in lanthanum-yttriumternary hydrides.* arXiv:2012.04787, 2020.
70. Prozorov, R., *Meissner-London Susceptibility of Superconducting Right Circular Cylinders in an Axial Magnetic Field.* Phys. Rev. Appl., 2021. **10**: p. 1.
71. Bean, C.P., *Magnetization of hard superconductors.* Phys. Rev. Lettr. , 1962. **8**: p. 250.
72. Prozorov, R., to be published, 2022.
73. Hong, F., et al., *Superconductivity at ~70 K in Tin Hydride SnHx under High Pressure.* arXiv:2101.02846, 2021.
74. Chen, W., et al., *High-Temperature Superconductivity in Cerium Superhydrides.* arXiv:2101.01315, 2021.